\documentclass[11pt,a4paper]{article}
\usepackage{jheppub}
\usepackage[T1]{fontenc}
\usepackage{lmodern}
\usepackage{microtype}
\usepackage{booktabs}
\usepackage{tabularx}
\usepackage{array}
\usepackage{bm}

\newcommand{\Tr}{\operatorname{Tr}}
\newcommand{\diag}{\operatorname{diag}}
\newcommand{\Hsp}{H_{\rm sp}}
\newcolumntype{Y}{>{\centering\arraybackslash}X}
\newcolumntype{L}{>{\raggedright\arraybackslash}X}

\title{\boldmath
Beyond a symmetry-restricted VEV ansatz:
transverse stability of an $SU(5)$ special-subgroup vacuum
}

\author{Hidetoshi Kawase}
\affiliation{CyberAgent, Inc., Shibuya, Tokyo 150--6121 Japan}
\emailAdd{kawase\_hidetoshi@cyberagent.co.jp}

\makeatletter
\gdef\@fpheader{ \vspace{1em} }
\makeatother

\abstract{
Extended Higgs sectors can realize multistage breaking chains in grand unified theories (GUTs), and the resulting intermediate gauge phases can be relevant to cosmological defects.
Restricting vacuum expectation values (VEVs) to configurations that preserve a chosen subgroup is an efficient way to identify candidate stationary points, while their stability against fluctuations outside this subspace is a separate question.
We study this question in a renormalizable $SU(5)$ GUT Higgs sector with an adjoint $24_H$ and a complex symmetric $15_H$, motivated by a special-subgroup realization of Langacker--Pi monopole erasure.
On the standard adjoint background, the $15_H$ VEVs that preserve the intermediate gauge group form a two-block family with independent color- and weak-block amplitudes.
Portal interactions can make these amplitudes unequal without changing the unbroken gauge group.
Two independent mixed quartic invariants become identical when evaluated on this two-block field space, although they contribute differently to color--weak fluctuations.
We derive the complete spectrum of these modes and identify two bounded-from-below parameter choices that induce the same potential on the two-block field space but have different curvatures in transverse directions.
The common stationary configuration is a local minimum in one case and a saddle with six negative physical modes in the other.
Direct Hessian calculations in the full 54-dimensional real scalar field space reproduce the analytic spectrum and verify both the minimum--saddle distinction and locally stable benchmark vacua.
}

\begin{document}
\maketitle
\flushbottom

\section{Introduction}
\label{sec:introduction}

Grand unified theories (GUTs) often employ extended Higgs sectors to realize multistage breaking chains, especially when several scalar representations participate in the breaking of a larger unified gauge group.
The resulting intermediate gauge phases can have important consequences for cosmological defects and early-universe symmetry breaking.
In the Georgi--Glashow model, the breaking
\begin{equation}
 SU(5)\longrightarrow
 H_{32}=S\bigl(U(3)\times U(2)\bigr)
\end{equation}
produces topologically stable GUT monopoles whose thermal abundance is unacceptable in a standard cosmological history \cite{Georgi:1974sy,tHooft:1974kcl,Polyakov:1974ek,Kibble:1976sj,Preskill:1979zi}.
The Langacker--Pi mechanism provides a possible dynamical resolution: a later symmetry breaking confines monopole flux into strings and can drive monopole--antimonopole annihilation \cite{Langacker:1980kd}.

Candidate intermediate phases are often identified by restricting the scalar vacuum expectation values (VEVs) to configurations that preserve a chosen subgroup.
We refer to such a restriction as a symmetry-restricted VEV ansatz.
It provides an efficient way to locate stationary configurations, but leaves two separate questions.
First, does the chosen ansatz span the full field subspace preserving the proposed subgroup?
Second, is a stationary configuration stable against fluctuations outside that subspace?
When the restricted configurations constitute the full fixed-point subspace of the chosen subgroup, their use for locating stationary points is closely related to the principle of symmetric criticality \cite{Palais:1979rca}.
This first-variation statement does not determine the Hessian in directions normal to the fixed subspace.
The present work addresses both questions in a concrete grand-unified Higgs sector.

We consider an $SU(5)$ model with an adjoint scalar $\Phi\in24_H$ and a complex symmetric tensor $S\in15_H$.
The high-scale breaking sequence of interest is
\begin{equation}
 SU(5)
 \xrightarrow{\langle\Phi\rangle}
 H_{32}=S\bigl(U(3)\times U(2)\bigr)
 \xrightarrow{\langle S\rangle}
 H_{\rm sp}=S\bigl(O(3)_C\times O(2)_L\bigr).
 \label{eq:high-scale-chain}
\end{equation}
This is the high-scale scalar core of the Hamada--Yamatsu realization of Langacker--Pi monopole erasure in $SU(5)$ \cite{Hamada:2026iht}.
Their construction provides the cosmological setting, while our analysis focuses on the vacuum geometry and local stability of the intermediate phase when the complete renormalizable $24_H$--$15_H$ portal sector is retained.

On the standard $3+2$ adjoint background preserving $H_{32}$, an $H_{\rm sp}$-preserving two-block family is
\begin{equation}
 S=\diag(s_C I_3,s_L I_2),
 \qquad s_Cs_L\neq0.
 \label{eq:hsp-family-intro}
\end{equation}
The isotropic representative $S\propto I_5$ adopted for the intermediate phase in the Hamada--Yamatsu construction corresponds to the special case $s_C=s_L$.
Portal interactions can shift a stationary point to unequal color- and weak-block amplitudes without changing the unbroken gauge group.
A second effect concerns fluctuations outside this two-block field space.
The renormalizable theory contains three independent mixed quartic invariants, two of which become identical on block-diagonal configurations but contribute differently to color--weak cross-block fluctuations.
The reduced potential therefore retains less coupling information than the full Hessian.
This illustrates a general mechanism: when two invariant interactions coincide on a symmetry-fixed field subspace, the reduced potential and its tangential Hessian depend only on a combination of their couplings, whereas the normal Hessian can still distinguish them.

We determine the stationary deformation of the two-block family and derive the complete color--weak cross-block spectrum, consisting of two sixfold physical families.
We identify two bounded-from-below parameter choices that induce the same potential on the two-block field space but different curvatures in transverse directions.
The common stationary configuration is a local minimum in one case and a saddle with six negative physical modes in the other.
Direct Hessian calculations in the full 54-dimensional real scalar field space reproduce the analytic spectrum and verify locally stable benchmark vacua.
We also compare the principal competing configurations and study their entry into the intermediate phase within a leading thermal-mass approximation.

Our results concern local vacuum geometry rather than a certified global phase diagram of the complete scalar theory.
The thermal analysis determines stationary-branch ordering within the leading-mass approximation; it does not address the later return transition, bubble nucleation, defect-network evolution, or gravitational-wave production.

The paper is organized as follows.
Section~\ref{sec:model} introduces the scalar potential and the portal-splitting identity.
Section~\ref{sec:branches} describes the relevant vacuum families and their stabilizers.
Section~\ref{sec:transverse} derives the stationary conditions and the complete cross-block spectrum.
Section~\ref{sec:zeroT} presents the zero-temperature comparisons and full-field validation.
Section~\ref{sec:thermal} discusses illustrative thermal-mass trajectories.

\section{Scalar potential and projected orbit variables}
\label{sec:model}

\subsection{Scalar potential}

We represent the adjoint scalar as a traceless Hermitian matrix,
\begin{equation}
 \Phi\in24_H,
 \qquad \Phi^\dagger=\Phi,
 \qquad \Tr\Phi=0,
\end{equation}
and the complex symmetric tensor as
\begin{equation}
 S\in15_H,
 \qquad S^T=S.
\end{equation}
Under $U\in SU(5)$,
\begin{equation}
 \Phi\longrightarrow U\Phi U^\dagger,
 \qquad
 S\longrightarrow USU^T.
 \label{eq:transformations}
\end{equation}
We use the kinetic convention
\begin{equation}
 \mathcal L_{\rm kin}
 =\frac12\Tr(D_\mu\Phi D^\mu\Phi)
 +\Tr(D_\mu S D^\mu S^\dagger).
 \label{eq:kinetic}
\end{equation}

The renormalizable scalar potential is
\begin{align}
 V(\Phi,S)={}&
 \mu_\Phi^2\Tr\Phi^2
 +a_\Phi\Tr\Phi^3
 +\lambda_1(\Tr\Phi^2)^2
 +\lambda_2\Tr\Phi^4
 \notag\\
 &+\mu_S^2\Tr(SS^\dagger)
 +\kappa_1[\Tr(SS^\dagger)]^2
 +\kappa_2\Tr[(SS^\dagger)^2]
 \notag\\
 &+\rho\Tr(\Phi SS^\dagger)
 +\alpha\Tr\Phi^2\Tr(SS^\dagger)
 \notag\\
 &+\beta\Tr(\Phi^2SS^\dagger)
 +\gamma\Tr(\Phi S\Phi^T S^\dagger).
 \label{eq:potential}
\end{align}

The scalar potential also possesses an accidental global symmetry,
\begin{equation}
 U(1)_S:\ S\longrightarrow e^{i\theta}S.
 \label{eq:U1S}
\end{equation}
When $S\neq0$, this symmetry gives one physical Goldstone mode.
In a complete model it may be lifted by Yukawa interactions or higher-dimensional operators; no such explicit breaking is included here.

\subsection{Three mixed quartic portals}

There are three independent quartic singlets containing two adjoint fields and one $S$--$S^\dagger$ pair.
Representation theory gives
\begin{equation}
 \operatorname{Sym}^2(\mathbf{24})
 =\mathbf{1}\oplus\mathbf{24}\oplus\mathbf{75}\oplus\mathbf{200},
 \qquad
 \mathbf{15}\otimes\overline{\mathbf{15}}
 =\mathbf{1}\oplus\mathbf{24}\oplus\mathbf{200},
 \label{eq:rep-decomp}
\end{equation}
Thus, the common irreducible components correspond to the $\alpha$-, $\beta$-, and $\gamma$-type contractions in eq.~\eqref{eq:potential} \cite{Slansky:1981yr}.

The two orientation-dependent contractions obey the basis-independent identity
\begin{align}
 &\Tr(\Phi^2SS^\dagger)-\Tr(\Phi S\Phi^T S^\dagger)
 \notag\\
 &\hspace{12mm}=\frac12\Tr\!\left[(\Phi S-S\Phi^T)(\Phi S-S\Phi^T)^\dagger\right]\ge0.
 \label{eq:pminusq-raw}
\end{align}
In a basis in which $\Phi=\diag(\phi_1,\ldots,\phi_5)$, the right-hand side becomes $\frac12\sum_{i,j}(\phi_i-\phi_j)^2|S_{ij}|^2$.
The two contractions therefore coincide whenever $S$ does not mix distinct eigenspaces of $\Phi$.
Consequently, a diagonal ansatz depends on $\beta$ and $\gamma$ only through $\beta+\gamma$, whereas color--weak off-diagonal fluctuations distinguish them.
The diagonal treatment in ref.~\cite{Hamada:2026iht} naturally captures the corresponding effective combination; the present analysis completes it in the normal directions.

\subsection{Projected orbit variables}

We separate the radial variables,
\begin{equation}
 r_\Phi^2=\Tr\Phi^2,
 \qquad
 r_S^2=\Tr(SS^\dagger),
\end{equation}
and introduce the dimensionless orbit variables
\begin{align}
 k&=\frac{\Tr\Phi^3}{(\Tr\Phi^2)^{3/2}},
 &
 \ell&=\frac{\Tr(\Phi SS^\dagger)}{\sqrt{\Tr\Phi^2}\,\Tr(SS^\dagger)},
 \notag\\
 u&=\frac{\Tr\Phi^4}{(\Tr\Phi^2)^2},
 &
 v&=\frac{\Tr[(SS^\dagger)^2]}{[\Tr(SS^\dagger)]^2},
 \notag\\
 p&=\frac{\Tr(\Phi^2SS^\dagger)}{\Tr\Phi^2\Tr(SS^\dagger)},
 &
 q&=\frac{\Tr(\Phi S\Phi^T S^\dagger)}{\Tr\Phi^2\Tr(SS^\dagger)}
 \label{eq:orbit-variables}
\end{align}
Invariant-polynomial and orbit-space descriptions of Higgs-potential minima and residual symmetries have a long history \cite{Abud:1981tf,Kim:1981xu,Abud:1983id,Sartori:1981zj,Frautschi:1981jh,Talamini:2006cf}.
Modern vacuum-stability analyses often combine gauge-orbit variables with copositivity methods \cite{Kannike:2016fmd,Chauhan:2019fji}.
In terms of these variables, the scalar potential can be organized by homogeneous degree as
\begin{equation}
 V=V_2+V_3+V_4,
 \label{eq:potential-by-degree}
\end{equation}
where
\begin{align}
 V_2&=\mu_\Phi^2r_\Phi^2+\mu_S^2r_S^2,
 \label{eq:V2}
 \\
 V_3&=a_\Phi k r_\Phi^3+\rho\,\ell\,r_\Phi r_S^2,
 \label{eq:V3}
 \\
 V_4&=r_\Phi^4A(u)+r_S^4B(v)+r_\Phi^2r_S^2C(p,q).
 \label{eq:V4}
\end{align}
with
\begin{equation}
 A(u)=\lambda_1+\lambda_2u,
 \qquad
 B(v)=\kappa_1+\kappa_2v,
 \qquad
 C(p,q)=\alpha+\beta p+\gamma q.
 \label{eq:ABC}
\end{equation}
The quartic form controls the large-field behavior except along
quartic-flat directions, where the lower-degree terms must also be checked.

For a traceless Hermitian $5\times5$ matrix and a complex symmetric matrix, respectively,
\begin{equation}
 \frac{7}{30}\le u\le\frac{13}{20},
 \qquad
 \frac15\le v\le1.
 \label{eq:uv-ranges}
\end{equation}
The lower and upper endpoints of the $u$ interval are attained by the $D_{32}$ and $D_{41}$ adjoint patterns, while $v=1/5$ corresponds to equal Takagi singular values and $v=1$ to a rank-one $S$.
These ranges are derived in appendix~\ref{app:orbit}.

\subsection{Basic quartic-stability conditions}

At each allowed orbit point, non-negativity of the quartic form is a copositivity problem in the nonnegative radial variables \cite{Kannike:2012pe,Chakrabortty:2013mha,Kannike:2016fmd}.
Gauge-orbit variables can further simplify this problem in theories with nontrivial scalar multiplets \cite{Chauhan:2019fji}.

Equation~\eqref{eq:uv-ranges} implies the pure-sector conditions
\begin{equation}
 \lambda_1+\frac{7}{30}\lambda_2\ge0,
 \qquad
 \lambda_1+\frac{13}{20}\lambda_2\ge0,
 \label{eq:pure-phi-bfb}
\end{equation}
\begin{equation}
 \kappa_1+\frac15\kappa_2\ge0,
 \qquad
 \kappa_1+\kappa_2\ge0.
 \label{eq:pure-s-bfb}
\end{equation}
At fixed orbit point, let $X=r_\Phi^2\ge0$ and $Y=r_S^2\ge0$. Then
\begin{equation}
 V_4=AX^2+BY^2+CXY
 =(\sqrt A X-\sqrt B Y)^2+(C+2\sqrt{AB})XY.
\end{equation}
The mixed-direction condition is therefore
\begin{equation}
 C(p,q)+2\sqrt{A(u)B(v)}\ge0,
 \label{eq:mixed-bfb}
\end{equation}
which is the copositivity criterion for this two-variable biquadratic form \cite{Kannike:2012pe,Chakrabortty:2013mha,Kannike:2016fmd}.
Imposed over the full projected orbit space, eqs.~\eqref{eq:pure-phi-bfb}, \eqref{eq:pure-s-bfb}, and \eqref{eq:mixed-bfb} are necessary and sufficient for $V_4\ge0$.
If these inequalities are strict throughout the orbit space, $V_4$ is positive definite and the full tree-level potential is bounded from below; if they are saturated, $V_3$ and $V_2$ must also be checked along the corresponding quartic-flat directions.
We impose eqs.~\eqref{eq:pure-phi-bfb} and \eqref{eq:pure-s-bfb} analytically.
For the two benchmarks used below, a positive analytic lower bound on eq.~\eqref{eq:mixed-bfb} is obtained over the full projected orbit space; see section~\ref{sec:bfb-benchmarks}.

\section{Vacuum branches}
\label{sec:branches}

\subsection{Adjoint alignment and the $D_{32}$ direction}

The orbit-space analysis of renormalizable $SU(N)$ adjoint Higgs potentials and their symmetry-breaking extrema is classical \cite{Ruegg:1980gf,Kim:1981jj}.
For $S=0$, it is useful to separate the adjoint radius from its orbit data,
\begin{equation}
 V_\Phi
 =\mu_\Phi^2r_\Phi^2+a_\Phi k r_\Phi^3
 +(\lambda_1+\lambda_2u)r_\Phi^4.
 \label{eq:adjoint-potential}
\end{equation}
When $a_\Phi=0$, radial minimization at fixed $u$ gives
\begin{equation}
 V_{\Phi,\min}(u)=-\frac{(\mu_\Phi^2)^2}{4(\lambda_1+\lambda_2u)}
 \label{eq:adjoint-radial}
\end{equation}
for $\mu_\Phi^2<0$ and a positive quartic coefficient.
Thus, when $\lambda_2>0$, the lower endpoint $u=7/30$ is selected and the adjoint expectation value is aligned with
\begin{equation}
 \Phi=\phi D_{32},
 \qquad
 D_{32}=\frac{1}{\sqrt{30}}\diag(2,2,2,-3,-3).
 \label{eq:D32}
\end{equation}
For nonzero $a_\Phi$, the cubic invariant also varies over the adjoint orbit, so $\lambda_2>0$ alone no longer establishes the preferred eigenvalue partition.

At a nonzero stationary point on the $D_{32}$ branch, the adjoint fluctuations that split the degeneracies within the color and weak blocks transform as $(\mathbf 8,\mathbf 1)_0$ and $(\mathbf 1,\mathbf 3)_0$.
After using radial stationarity to eliminate $\mu_\Phi^2$, their curvatures are
\begin{align}
 m_8^2
 &=
 \frac{2}{3}\lambda_2\phi^2
 +\frac{15}{\sqrt{30}}a_\Phi\phi,
 \notag\\
 m_3^2
 &=
 \frac{8}{3}\lambda_2\phi^2
 -\frac{15}{\sqrt{30}}a_\Phi\phi.
 \label{eq:D32-angular-masses}
\end{align}
For the orientation in eq.~\eqref{eq:D32} with $\phi>0$, local angular stability therefore requires
\begin{equation}
 -\frac{2\sqrt{30}}{45}\lambda_2\phi
 <a_\Phi<
 \frac{8\sqrt{30}}{45}\lambda_2\phi.
 \label{eq:D32-angular-stability}
\end{equation}
For $a_\Phi=0$, this reduces to the condition $\lambda_2>0$. A pure-adjoint local minimum additionally requires a positive radial curvature, while global selection over other eigenvalue partitions, in particular the $D_{41}$ branch, requires a comparison of the radially minimized branch energies.

The mixed-sector analysis below is conditioned on a locally stable $D_{32}$ branch.
The numerical reference slice takes $a_\Phi=0$ and $\lambda_2=1$; the unrestricted calculation verifies that the solution remains aligned at $u=7/30$ after the $15_H$ condensate forms.

\subsection{The $H_{\rm sp}$-preserving family}

The background in eq.~\eqref{eq:D32} preserves
\begin{equation}
 H_{32}=S\bigl(U(3)\times U(2)\bigr).
\end{equation}
Using the residual transformations and the global phase in eq.~\eqref{eq:U1S}, we take the block expectation values to be real and write
\begin{equation}
 S=\diag(s_C I_3,s_L I_2),
 \qquad s_C>0,
 \quad s_L>0.
 \label{eq:Hsp-family}
\end{equation}
It is useful to introduce
\begin{equation}
 s^2=3s_C^2+2s_L^2,
 \qquad
 w=\frac{3s_C^2}{s^2},
 \qquad 0<w<1.
 \label{eq:w-def}
\end{equation}
The isotropic point $s_C=s_L$ corresponds to $w=3/5$, while $w\to1$ and $w\to0$ approach the color-block and weak-block endpoints.

On this family,
\begin{equation}
 v(w)=\frac15+\frac56\left(w-\frac35\right)^2,\qquad
 p(w)=q(w)=\frac{3}{10}-\frac{w}{6},\qquad
 \ell(w)=\frac{5w-3}{\sqrt{30}}.
 \label{eq:Hsp-orbit}
\end{equation}
Equations~\eqref{eq:V3}--\eqref{eq:ABC} and \eqref{eq:Hsp-orbit} show that $\kappa_2>0$ favors the isotropic point $s_C=s_L$, while $\rho$ and $\beta+\gamma$ can bias the stationary solution toward unequal block amplitudes.

\subsection{Stabilizer}

Let $U\in SU(5)$ preserve both backgrounds in eqs.~\eqref{eq:D32} and \eqref{eq:Hsp-family}.
The condition $U\Phi U^\dagger=\Phi$ requires
\begin{equation}
 U=\diag(A,B),
 \qquad
 A\in U(3),
 \quad B\in U(2),
 \quad \det A\det B=1.
 \label{eq:block-U}
\end{equation}
The second condition, $USU^T=S$, gives
\begin{equation}
 AA^T=I_3,
 \qquad
 BB^T=I_2,
 \label{eq:orthogonal-condition}
\end{equation}
provided $s_Cs_L\neq0$.
A unitary matrix satisfying eq.~\eqref{eq:orthogonal-condition} is real orthogonal.
The common stabilizer is therefore
\begin{equation}
 \begin{aligned}
 \Hsp
 &=S\bigl(O(3)\times O(2)\bigr)\\
 &=\left\{\diag(A,B)\,\middle|\,
 A\in O(3),\ B\in O(2),\ \det A\det B=1\right\}.
 \end{aligned}
 \label{eq:Hsp-stabilizer}
\end{equation}
with connected component
\begin{equation}
 \Hsp{}_0=SO(3)\times SO(2).
\end{equation}
The result is independent of $s_C/s_L$.
Portal-induced block distortion therefore leaves the unbroken gauge group unchanged as long as both blocks remain nonzero.
At either endpoint, one of the conditions in eq.~\eqref{eq:orthogonal-condition} is lost and the stabilizer is enlarged, so the endpoints belong to different vacuum branches.

Equation~\eqref{eq:Hsp-stabilizer} also fixes the number of broken gauge directions.
A generic $H_{\rm sp}$ background has
\begin{equation}
 24-\dim[SO(3)\times SO(2)]=20
\end{equation}
broken generators, which provides a useful check on the full scalar Hessian.

\subsection{Representative competitors}

The zero-temperature ordering is studied using the continuous $H_{\rm sp}$ family and the principal competing alignments listed in table~\ref{tab:branches}.
We refer to configurations in which $S$ has support only in the off-diagonal block connecting the color and weak eigenspaces of $D_{32}$ as color--weak (CW) configurations.

For $\kappa_2<0$, the pure-$S$ quartic favors the endpoint $v=1$, so rank-one tensors must also be included.
A normalized rank-one configuration can be written as
\begin{equation}
 S=s\,zz^T,
 \qquad z^\dagger z=1.
\end{equation}
The residual $U(3)\times U(2)$ symmetry of the $D_{32}$ background reduces its orientation to the color weight
\begin{equation}
 t\equiv\sum_{a=1}^3|z_a|^2\in[0,1].
\end{equation}
Its orbit variables are
\begin{equation}
 v=1,
 \qquad
 p=\frac{3}{10}-\frac{t}{6},
 \qquad
 q=\frac{(5t-3)^2}{30},
 \qquad
 \ell=\frac{5t-3}{\sqrt{30}}.
 \label{eq:rank-one-orbit}
\end{equation}
The endpoints $t=1$ and $t=0$ are color- and weak-aligned rank-one configurations, while $0<t<1$ describes a mixed orientation.

\begin{table}[t]
\centering
\begin{tabular}{llccc}
\toprule
branch & schematic $S$ & stabilizer $H_{\Phi,S}$ & $(p,q)$ & $v$ \\
\midrule
$H_{\rm sp}$ diagonal family
& $\diag(s_C I_3,s_L I_2)$
& $S\!\left(O(3)\times O(2)\right)$
& $(p(w),p(w))$
& $v(w)$ \\
color-block branch
& $\diag(xI_3,0_2)$
& $S\!\left(O(3)\times U(2)\right)$
& $(2/15,2/15)$
& $1/3$ \\
weak-block branch
& $\diag(0_3,yI_2)$
& $S\!\left(U(3)\times O(2)\right)$
& $(3/10,3/10)$
& $1/2$ \\
CW balanced branch
& $\left(\begin{smallmatrix}0&X\\X^T&0\end{smallmatrix}\right)$
& $U(2)_{\rm CW}$
& $(13/60,-1/5)$
& $1/4$ \\
rank-one family
& $s\,zz^T$
& $\mathbb Z_2\times U(2)$
& $(p(t),q(t))$
& $1$ \\
\bottomrule
\end{tabular}
\caption{
Candidate configurations on the $\Phi=\phi D_{32}$ background, their common stabilizers $H_{\Phi,S}\subset SU(5)$, and the orbit data used in the restricted zero-temperature and thermal comparisons.
Here $X$ is a $3\times2$ CW block.
The CW balanced branch is defined by $X$ having two equal nonzero singular values.
For the rank-one family, the stabilizer is $\mathbb Z_2\times U(2)$ for $0<t<1$ and is enhanced at the two endpoints.
The functions $p(w)$ and $v(w)$ are defined in eq.~\eqref{eq:Hsp-orbit}; the rank-one functions $p(t)$ and $q(t)$ are given in eq.~\eqref{eq:rank-one-orbit}.
}
\label{tab:branches}
\end{table}

For the CW balanced branch, $q=-1/5$ follows from the two-block off-diagonal relation, $p=13/60$ from the color and weak eigenvalue squares of $D_{32}$, and $v=1/4$ from two equal nonzero singular values.
The derivation is given in appendix~\ref{app:CW}.
Positive $\gamma$ lowers the $q<0$ CW candidate, while the cubic portal $\rho$ can favor the color- or weak-block endpoints.
The ordering of the $H_{\rm sp}$ family is therefore determined by the combined effects of diagonal distortion, off-diagonal alignment, and radial minimization.

\section{Portal-driven deformation and transverse stability}
\label{sec:transverse}

\subsection{Stationary conditions on the $H_{\rm sp}$ family}

Substituting eqs.~\eqref{eq:D32} and \eqref{eq:Hsp-family} into the potential gives
\begin{align}
 V_{\rm diag}={}&
 \mu_\Phi^2\phi^2-\frac{a_\Phi}{\sqrt{30}}\phi^3
 +\left(\lambda_1+\frac{7}{30}\lambda_2\right)\phi^4
 +\mu_S^2(3s_C^2+2s_L^2)
 \notag\\
 &+\kappa_1(3s_C^2+2s_L^2)^2
 +\kappa_2(3s_C^4+2s_L^4)
 \notag\\
 &+\frac{6\rho}{\sqrt{30}}\phi(s_C^2-s_L^2)
 +\alpha\phi^2(3s_C^2+2s_L^2)
 \notag\\
 &+\frac{\beta+\gamma}{5}\phi^2(2s_C^2+3s_L^2).
 \label{eq:Vdiag}
\end{align}
The three stationary-point equations are
\begin{align}
 0={}&\mu_\Phi^2\phi-\frac{3a_\Phi}{2\sqrt{30}}\phi^2
 +2\left(\lambda_1+\frac{7}{30}\lambda_2\right)\phi^3
 +\frac{3\rho}{\sqrt{30}}(s_C^2-s_L^2)
 \notag\\
 &+\alpha\phi(3s_C^2+2s_L^2)
 +\frac{\beta+\gamma}{5}\phi(2s_C^2+3s_L^2),
 \label{eq:stat-phi}
\end{align}
\begin{equation}
 0=\mu_S^2+2\kappa_1(3s_C^2+2s_L^2)+2\kappa_2s_C^2
 +\frac{2\rho}{\sqrt{30}}\phi+\alpha\phi^2
 +\frac{2}{15}(\beta+\gamma)\phi^2,
 \label{eq:stat-sc}
\end{equation}
\begin{equation}
 0=\mu_S^2+2\kappa_1(3s_C^2+2s_L^2)+2\kappa_2s_L^2
 -\frac{3\rho}{\sqrt{30}}\phi+\alpha\phi^2
 +\frac{3}{10}(\beta+\gamma)\phi^2.
 \label{eq:stat-sl}
\end{equation}
Subtracting eqs.~\eqref{eq:stat-sc} and \eqref{eq:stat-sl} yields the block-balance condition
\begin{equation}
 12\kappa_2(s_C^2-s_L^2)+\sqrt{30}\rho\phi-(\beta+\gamma)\phi^2=0.
 \label{eq:block-balance}
\end{equation}
It makes explicit that $\kappa_2$ tends to equalize the two blocks, while the cubic portal and the diagonal quartic combination displace their ratio.
The isotropic point is stationary only on a codimension-one relation among these couplings.
Notice that $a_\Phi$ drops out of eq.~\eqref{eq:block-balance}; it affects the adjoint radial equation but not the relative block balance.

\subsection{Two sixfold color--weak relative-orientation families}

For each color index $a=1,2,3$ and weak index $\alpha=4,5$, the off-diagonal fluctuations of $\Phi$ and $S$ decompose into two inequivalent real pair blocks.
Each block contains one broken gauge tangent and one physical relative orientation.
A convenient matrix-unit basis, the two gauge tangents, and the kinetic metric are given in appendix~\ref{app:transverse}.
Since the same two blocks occur for all $3\times2=6$ color--weak pairs, the complete physical cross-block spectrum consists of two sixfold families.

We label the two real structures by $\eta=\pm1$ and define
\begin{equation}
 \Xi_\eta\equiv12\kappa_2(s_C-\eta s_L)^2-5\gamma\phi^2.
 \label{eq:Xi-eta}
\end{equation}
Using the full stationary-point conditions in eqs.~\eqref{eq:stat-phi}--\eqref{eq:stat-sl}, reducing each pair Hessian, and removing its gauge direction gives
\begin{equation}
 m_{\perp,\eta}^2=
 \left[\frac{1}{12}+\frac{(s_C+\eta s_L)^2}{5\phi^2}\right]
 \Xi_\eta,
 \qquad \eta=\pm1.
 \label{eq:transverse-masses}
\end{equation}
The $\eta=+1$ family is
\begin{equation}
 m_{\perp,+}^2=
 \left[\frac{1}{12}+\frac{(s_C+s_L)^2}{5\phi^2}\right]
 \left[12\kappa_2(s_C-s_L)^2-5\gamma\phi^2\right],
 \label{eq:transverse-plus}
\end{equation}
which is the softer family near the isotropic point.
The complementary family is
\begin{equation}
 m_{\perp,-}^2=
 \left[\frac{1}{12}+\frac{(s_C-s_L)^2}{5\phi^2}\right]
 \left[12\kappa_2(s_C+s_L)^2-5\gamma\phi^2\right].
 \label{eq:transverse-minus}
\end{equation}
The first factor in each expression is manifestly positive.
The terms proportional to $\kappa_2$ are the restoring forces associated with the two block combinations, while the $\gamma$ term is the crossed-portal contribution that is invisible as an independent coupling on the diagonal ansatz.
In particular,
\begin{equation}
 \kappa_2>0,
 \qquad
 \gamma<0
 \label{eq:simple-transverse-sufficient}
\end{equation}
is sufficient for both sixfold families to be non-tachyonic.

Equation~\eqref{eq:transverse-masses} addresses the complete color--weak cross-block sector.
Radial, block-ratio, and block-internal stability must still be checked separately.
We therefore use the full 54-dimensional Hessian in section~\ref{sec:fullfield} to establish local minima at representative points.

\subsection{A minimum-to-saddle boundary at fixed reduced potential}
\label{sec:transverse-boundary}

The reduced potential in eq.~\eqref{eq:Vdiag} and its stationarity conditions depend on $\beta$ and $\gamma$ only through
\begin{equation}
 \sigma\equiv\beta+\gamma.
 \label{eq:sigma-definition}
\end{equation}
Consequently, the fixed-$\sigma$ deformation
\begin{equation}
 \beta\longrightarrow\beta-\delta,
 \qquad
 \gamma\longrightarrow\gamma+\delta
 \label{eq:fixed-sigma-deformation}
\end{equation}
leaves the reduced potential, the two-block stationary solution $(\phi,s_C,s_L)$, its stationary-point energy, and the Hessian restricted to the two-block field space unchanged.

The complete cross-block spectrum does not share this degeneracy, because eq.~\eqref{eq:transverse-masses} retains a separate dependence on $\gamma$.
Since the first factor in each mass is positive, the two sixfold families change sign at
\begin{equation}
 \gamma_{{\rm crit},\eta}
 =
 \frac{12\kappa_2(s_C-\eta s_L)^2}{5\phi^2},
 \qquad
 \eta=\pm1.
 \label{eq:gamma-critical}
\end{equation}
For $\kappa_2>0$ near the isotropic point, $\gamma_{{\rm crit},+}\propto(s_C-s_L)^2$ is parametrically smaller than $\gamma_{{\rm crit},-}\propto(s_C+s_L)^2$.
The $\eta=+1$ family is therefore the first to become tachyonic as $\gamma$ is increased at fixed $\sigma$.

Provided the remaining physical eigenvalues stay positive, crossing $\gamma_{{\rm crit},+}$ changes the stationary configuration from a local minimum to a saddle with six negative physical modes, without changing any quantity visible in the reduced potential.
An explicit bounded-from-below pair realizing this distinction at $\sigma=0.02$ is given in appendix~\ref{app:stable-saddle-pair}.
This defines a local-stability boundary in parameter space.
It should not by itself be interpreted as a thermodynamic phase boundary, which would require following the branch emerging from the zero mode and comparing the competing branch energies.

\section{Zero-temperature benchmarks and full-field validation}
\label{sec:zeroT}

\subsection{Restricted candidate ordering and benchmark selection}
\label{sec:restricted-ordering}

We express dimensionful quantities in units of a fixed scale $M$,
\begin{equation}
 \widehat a_\Phi=\frac{a_\Phi}{M},
 \qquad
 \widehat\rho=\frac{\rho}{M},
 \qquad
 \widehat\mu_i^2=\frac{\mu_i^2}{M^2},
\end{equation}
and use the reference slice
\begin{equation}
 \widehat a_\Phi=0,
 \qquad
 \widehat\mu_\Phi^2=\widehat\mu_S^2=-1.5,
 \qquad
 \lambda_1=\lambda_2=\kappa_1=\kappa_2=1,
 \qquad
 \alpha=0.1.
 \label{eq:benchmark-slice}
\end{equation}
The choice $\widehat a_\Phi=0$ belongs only to this numerical reference slice.
It removes one dimensionful benchmark parameter and makes the fixed-$\sigma$ comparison especially transparent; the analytic block-balance relation and transverse spectrum above were obtained for arbitrary $a_\Phi$.

The analytic conditions in section~\ref{sec:transverse} determine local stability but do not show whether the portal-distorted branch is energetically competitive.
We therefore compare it with the principal candidate families in table~\ref{tab:branches} and use the resulting restricted ordering only to select representative points for unrestricted validation.
At each coupling point, the radial stationarity equations are solved on every family; $w$ is varied continuously on the $H_{\rm sp}$ family and the color weight $t$ on the rank-one family.
We retain minima with respect to the radial amplitudes and the available alignment coordinate, and require both cross-block masses in eq.~\eqref{eq:transverse-masses} to be non-tachyonic on the $H_{\rm sp}$ family.
The map displays the lowest-energy member of this comparison set and is not a phase diagram of the unrestricted 54-dimensional theory.

For the candidate-ordering scan we further set $\beta=0.1$.
For negative $\gamma$ and moderate $|\widehat\rho|$, the $H_{\rm sp}$ family is the lowest candidate over a broad region.
Positive $\gamma$ lowers the CW candidate, whereas increasing $|\widehat\rho|$ eventually drives the solution toward a block endpoint.
In the negative-$\kappa_2$ strip, the rank-one family becomes the lowest represented candidate over most of the displayed range, as expected from the preference for the maximal value $v=1$.

\begin{figure}[t]
 \centering
 \includegraphics[width=\linewidth]{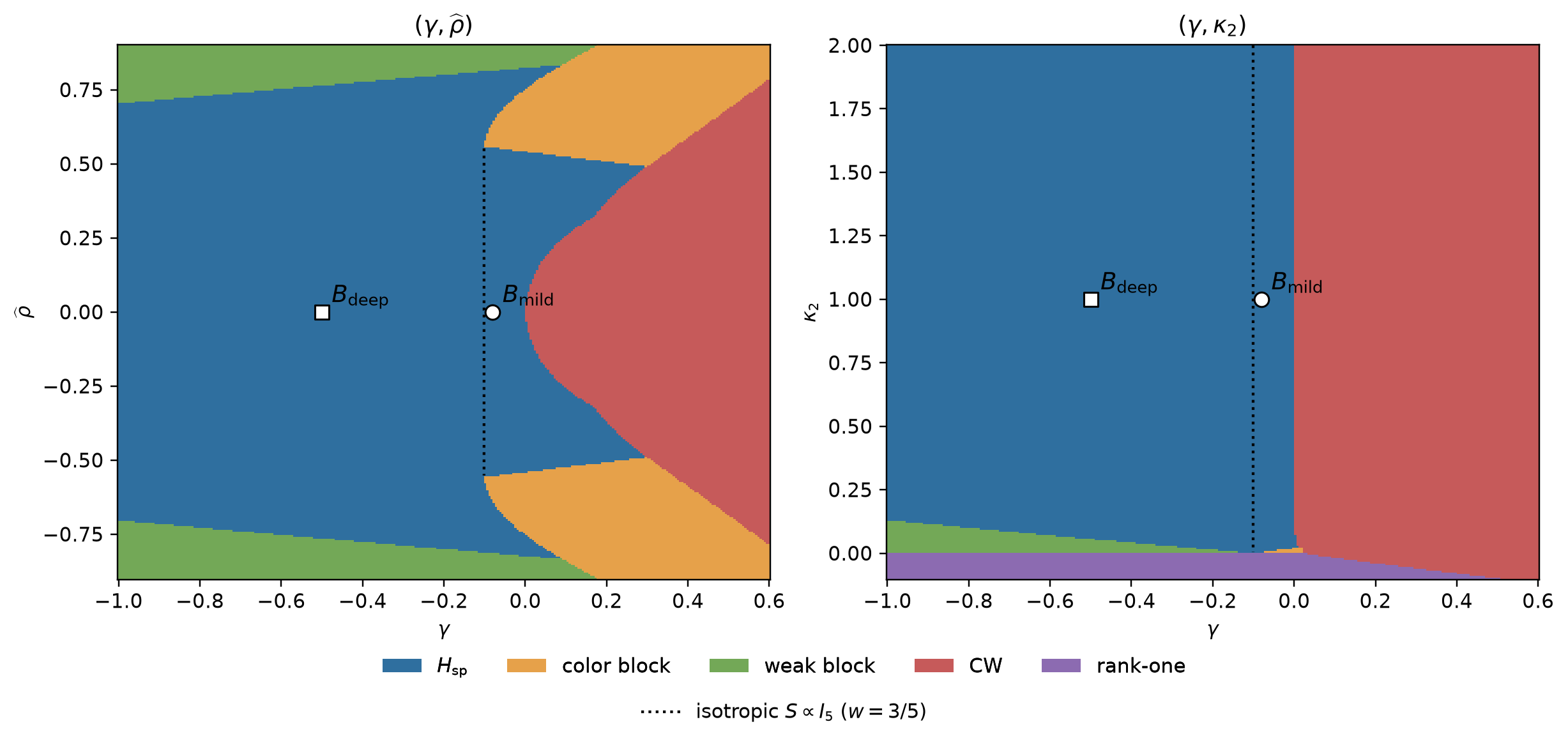}
 \caption{
 Restricted zero-temperature ordering of candidate branches.
 Left: the $(\gamma,\widehat\rho)$ plane with $-0.9\le\widehat\rho\le0.9$ and $\kappa_2=1$.
 Right: the $(\gamma,\kappa_2)$ plane with $-0.1\le\kappa_2\le2.0$ and $\widehat\rho=0$.
 At each point, the continuous $H_{\rm sp}$ and rank-one families are minimized together with the color-block, weak-block, and color--weak (CW) candidates, and the lowest-energy retained candidate is displayed.
 The markers indicate the two benchmarks subjected to full-field validation.
 The exactly isotropic condition $w=3/5$ is a codimension-one contour within the $H_{\rm sp}$ region, not a distinct vacuum stratum or local-stability boundary; its two sides correspond to $s_C>s_L$ and $s_C<s_L$.
 }
 \label{fig:restricted-ordering}
\end{figure}

Figure~\ref{fig:restricted-ordering} is not a certified phase diagram of the full 54-dimensional field space.
Its purpose is to organize the principal competing alignments and select representative $H_{\rm sp}$ points for unrestricted validation.

\subsection{Unrestricted local-stability test}
\label{sec:fullfield}

For the full-field calculation, the fields are expanded in 54 real coordinates canonically normalized with respect to eq.~\eqref{eq:kinetic}.
The adjoint is written as
\begin{equation}
 \Phi=\sum_{A=1}^{24}x_A B_A,
 \qquad
 \Tr(B_AB_B)=\delta_{AB},
 \label{eq:adjoint-basis}
\end{equation}
where $B_A$ is a traceless Hermitian basis.
For the complex symmetric tensor, we use
\begin{equation}
 S_{ii}=\frac{y_{ii}+iz_{ii}}{\sqrt2},
 \qquad
 S_{ij}=S_{ji}=\frac{y_{ij}+iz_{ij}}{2}
 \quad (i<j).
 \label{eq:symmetric-coordinates}
\end{equation}
With these factors,
\begin{equation}
 \mathcal L_{\rm kin}^{(2)}=\frac12\sum_{I=1}^{54}(\partial_\mu\varphi_I)^2.
\end{equation}

The matrix potential in eq.~\eqref{eq:potential} is evaluated directly, and its gradient and Hessian are generated by double-precision automatic differentiation.
The analytic pair block in appendix~\ref{app:transverse} is not used in constructing the full Hessian, so the comparison with eq.~\eqref{eq:transverse-masses} is independent.
Restricted minima are used as initial conditions for unconstrained quasi-Newton minimization in the 54-dimensional space.
We monitor the dimensionless stationarity residual
\begin{equation}
 \epsilon_{\rm stat}=\frac{\|\nabla V\|_2}{M^3}
 \label{eq:stationarity-residual}
\end{equation}
and require $\epsilon_{\rm stat}<10^{-10}$ for the reported benchmark solutions.

For each Hermitian $SU(5)$ generator $T_A$, we construct the gauge tangent
\begin{equation}
 \delta_A\Phi=i[T_A,\Phi],
 \qquad
 \delta_A S=i(T_AS+ST_A^T),
 \label{eq:gauge-tangents-full}
\end{equation}
and add the $U(1)_S$ tangent $(\delta\Phi,\delta S)=(0,iS)$.
A generic $H_{\rm sp}$ background has $\dim SU(5)-\dim H_{\rm sp}=24-4=20$ broken gauge generators.
Together with the Goldstone mode associated with the spontaneous breaking of the accidental $U(1)_S$, the ungauge-fixed scalar Hessian is therefore expected to contain 21 zero modes.
The twelve color--weak relative-orientation modes in eq.~\eqref{eq:transverse-masses} are not symmetry zero modes and are generically massive.
We use the zero-mode count as a diagnostic of the full-field Hessian.
The independent tangent vectors are orthonormalized by singular-value decomposition, and the Hessian is projected onto their orthogonal complement to obtain the physical spectrum.
A tolerance of $10^{-9}M^2$ is used for zero modes, together with an explicit check that the Hessian annihilates the symmetry tangents.
Branch crossings are refined by solving for zeros of the energy differences between coexisting branches; when a branch terminates, its existence boundary is determined by bisection.
These procedures refine the local spectra and the ordering of the represented branches but do not constitute a proof of global minimality in the full field space.

We now apply this setup to two portal-distorted benchmarks selected from the restricted comparison above.
We use two $H_{\rm sp}$ benchmarks,
\begin{equation}
 \begin{aligned}
  B_{\rm mild}:&\quad \beta=0.1,\quad \gamma=-0.08,\quad \widehat\rho=0,\\
  B_{\rm deep}:&\quad \beta=0.1,\quad \gamma=-0.5,\quad \widehat\rho=0.
 \end{aligned}
 \label{eq:benchmarks}
\end{equation}
The refined full-field results are summarized in table~\ref{tab:benchmark-validation}.
We denote by $w_*$ the value of the alignment parameter at the refined stationary point and quote both analytic cross-block masses from eq.~\eqref{eq:transverse-masses}.

\begin{table}[t]
\centering
\begin{tabular}{lccccc}
\toprule
benchmark & $w_*$ & $m_{\perp,+}^2/M^2$ & $m_{\perp,-}^2/M^2$ & $\epsilon_{\rm stat}$ & $(n_-,n_0)$ \\
\midrule
$B_{\rm mild}$ & $0.6019$ & $0.05779$ & $0.4988$ & $2.0\times10^{-13}$ & $(0,21)$ \\
$B_{\rm deep}$ & $0.5610$ & $0.3798$ & $0.6310$ & $1.0\times10^{-13}$ & $(0,21)$ \\
\bottomrule
\end{tabular}
\caption{
Full-field validation of the two $H_{\rm sp}$ benchmarks.
The quantities $n_-$ and $n_0$ are the number of negative physical Hessian eigenvalues and the total number of zero modes, respectively; $\epsilon_{\rm stat}$ is defined in eq.~\eqref{eq:stationarity-residual}.
}
\label{tab:benchmark-validation}
\end{table}

Both solutions lie in the interior of the portal-distorted family, with $0<w_*<1$ and $w_*\neq3/5$.
After unrestricted refinement,
\begin{equation}
 u_*=\frac{\Tr\Phi^4}{(\Tr\Phi^2)^2}=\frac{7}{30}
\end{equation}
within numerical precision, showing that the adjoint remains $D_{32}$-aligned after the $15_H$ condensate forms.
The physical Hessian has no negative eigenvalues and contains the expected 21 zero modes.
The structured configurations in table~\ref{tab:branches}, together with nearby random perturbations, were also used as initial conditions; no lower stationary solution was found within this search.
This is numerical evidence, not a global-minimum certificate.

\begin{figure}[t]
 \centering
 \includegraphics[width=0.96\linewidth]{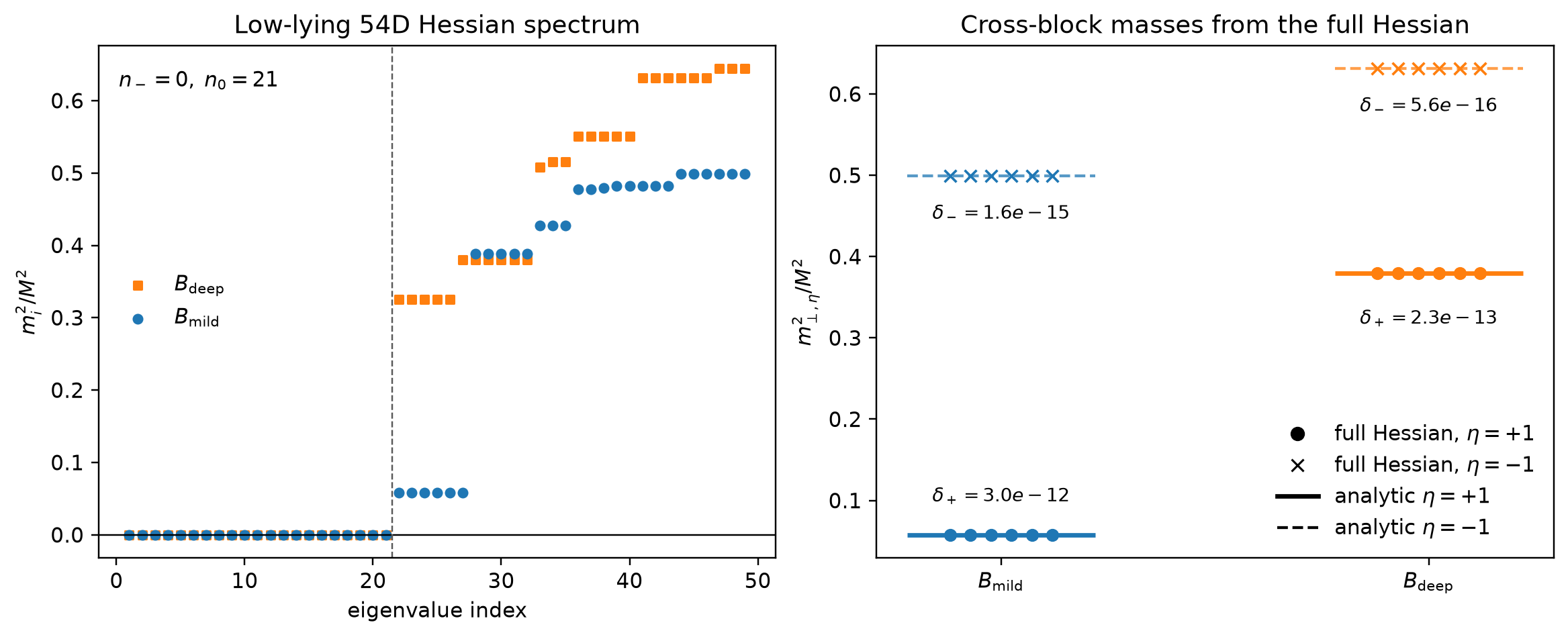}
 \caption{
 Full-field validation of the two $H_{\rm sp}$ benchmarks.
 Left: the low-lying spectrum of the canonically normalized 54-dimensional scalar Hessian.
 The positive physical spectrum begins after the 21 symmetry zero modes and contains both sixfold cross-block clusters.
 Right: the softer $\eta=+1$ relative-orientation family extracted from the full Hessian (points), compared with the analytic result in eq.~\eqref{eq:transverse-plus} (horizontal lines).
 }
 \label{fig:fullfield-validation}
\end{figure}

The twelve cross-block relative-orientation eigenvalues occur in two sixfold clusters and agree with eq.~\eqref{eq:transverse-masses} within numerical precision.
The right panel of figure~\ref{fig:fullfield-validation} displays the softer $\eta=+1$ family; the complementary sixfold cluster is visible in the full spectrum in the left panel and is listed in table~\ref{tab:benchmark-validation}.

\subsection{Quartic stability of the benchmarks}
\label{sec:bfb-benchmarks}

On the slice in eq.~\eqref{eq:benchmark-slice},
\begin{equation}
 A_{\min}=\lambda_1+\frac{7}{30}\lambda_2=\frac{37}{30},
 \qquad
 B_{\min}=\kappa_1+\frac15\kappa_2=\frac65.
 \label{eq:AminBmin}
\end{equation}
For an arbitrary field configuration,
\begin{equation}
 p-q\ge0,
 \qquad
 0\le p\le\frac45.
 \label{eq:p-bounds}
\end{equation}
The first inequality follows from eq.~\eqref{eq:pminusq-raw}.
To obtain the second, diagonalize $\Phi$ and normalize $\Tr\Phi^2=1$.
Writing
\begin{equation}
 w_i=\frac{(SS^\dagger)_{ii}}{\Tr(SS^\dagger)},
 \qquad
 w_i\geq0,
 \qquad
 \sum_i w_i=1,
\end{equation}
one has
\begin{equation}
 p=\sum_i w_i\phi_i^2
 \leq \max_i\phi_i^2.
\end{equation}
If $|x|$ is the largest absolute eigenvalue, tracelessness and Cauchy--Schwarz inequalities imply
\begin{equation}
 1=x^2+\sum_{j\neq i}\phi_j^2
 \geq x^2+\frac{x^2}{4},
\end{equation}
so $x^2\leq4/5$ and hence $0\leq p\leq4/5$.

For $\beta\ge0$ and $\gamma\le0$,
\begin{equation}
 C(p,q)\ge\alpha+\min\left[0,\frac45(\beta+\gamma)\right].
\end{equation}
Thus the left-hand side of eq.~\eqref{eq:mixed-bfb} obeys the global lower bound
\begin{equation}
 F\ge F_{\rm lb}
 \equiv\alpha+\min\left[0,\frac45(\beta+\gamma)\right]
 +2\sqrt{A_{\min}B_{\min}}.
 \label{eq:BFB-lower-bound}
\end{equation}
Numerically,
\begin{equation}
 F_{\rm lb}=2.533\dots\quad(B_{\rm mild}),
 \qquad
 F_{\rm lb}=2.213\dots\quad(B_{\rm deep}).
\end{equation}
Since $A_{\min}>0$, $B_{\min}>0$, and $F_{\rm lb}>0$, $V_4$ is strictly positive away from the origin at both benchmark points.
The full tree-level potential is therefore bounded from below.

\section{Thermal-mass branch ordering}
\label{sec:thermal}

\subsection{Thermal-mass deformation}

To model the leading high-temperature scalar self-energies, we use
\begin{equation}
 \mu_\Phi^2(T)=\mu_{\Phi,0}^2+c_\Phi T^2,
 \qquad
 \mu_S^2(T)=\mu_{S,0}^2+c_ST^2
 \label{eq:thermal-masses}
\end{equation}
\cite{Dolan:1973qd}.
At fixed quartic couplings, cooling traces a straight line in the $(\mu_\Phi^2,\mu_S^2)$ plane.
The analysis in this section is a diagnostic based on stationary-branch energies, branch existence, and their crossings.

We choose
\begin{equation}
 c_\Phi=0.4,
 \qquad
 c_S=0.6.
 \label{eq:thermal-coefficients}
\end{equation}
These are illustrative positive coefficients rather than one-loop values derived from a complete $SU(5)$ model.
For $\widehat\mu_{\Phi,0}^2=\widehat\mu_{S,0}^2=-1.5$, they make the adjoint instability occur before the $15_H$ instability.
Ignoring portals, the corresponding zero-crossing scales are
\begin{equation}
 \frac{T_{\Phi,0}}{M}=\sqrt{\frac{1.5}{0.4}}\simeq1.936,
 \qquad
 \frac{T_{S,0}}{M}=\sqrt{\frac{1.5}{0.6}}\simeq1.581.
\end{equation}

\subsection{Color--weak orientation at the onset of condensation}

For the two benchmarks, $\widehat\rho=0$. On a $D_{32}$ background, the orientation-dependent quadratic coefficient for an infinitesimal diagonal $S$ condensate is, up to a common factor,
\begin{equation}
 C_{\rm diag}(p)=\alpha+\sigma p,
 \qquad
 \sigma=\beta+\gamma.
 \label{eq:Cdiag}
\end{equation}
We denote the color-block and weak-block directions by CC and WW, respectively.
Together with the isotropic direction, their orbit values are
\begin{equation}
 p_{\rm CC}=\frac{2}{15},
 \qquad
 p_{\rm iso}=\frac15,
 \qquad
 p_{\rm WW}=\frac{3}{10}.
\end{equation}
Thus,
\begin{equation}
 C_{\rm CC}-C_{\rm iso}=-\frac{\sigma}{15},
 \qquad
 C_{\rm WW}-C_{\rm iso}=\frac{\sigma}{10}.
 \label{eq:onset-splitting}
\end{equation}
Thus the color-block direction is softest when $\sigma>0$, while the weak-block direction is softest when $\sigma<0$.
This criterion applies near the onset of the $S$ condensate.
At finite amplitude, quartic terms, including positive $\kappa_2$, can favor an interior solution with both blocks nonzero.

\subsection{Branch crossings and local stability}

At each temperature, we follow the symmetric, $D_{32}$, $S$-only, continuous $H_{\rm sp}$, color-block, weak-block, CW, and rank-one branches.
When two branches coexist and exchange order, the crossing temperature is obtained from the zero of their energy difference.
A branch endpoint is located by bisection when one solution ceases to exist.
The crossing temperatures in table~\ref{tab:thermal-crossings} are listed from high to low temperature.
Let $T_{\rm in}$ denote the crossing from $D_{32}$ into the intermediate color- or weak-block branch, and let $T_{\rm out}$ denote the crossing from that branch into $H_{\rm sp}$.
We define the width of the intermediate interval as
\begin{equation}
 \Delta T\equiv T_{\rm in}-T_{\rm out}>0.
 \label{eq:intermediate-width}
\end{equation}

\begin{table}[t]
\centering
\begin{tabularx}{\textwidth}{lXcc}
\toprule
benchmark & branch sequence & crossings $T/M$ & $\Delta T/M$ \\
\midrule
$B_{\rm mild}$ & symmetric $\to D_{32}\to$ CC $\to H_{\rm sp}$ & $1.936,\ 1.570,\ 1.568$ & $1.52\times10^{-3}$ \\
$B_{\rm deep}$ & symmetric $\to D_{32}\to$ WW $\to H_{\rm sp}$ & $1.936,\ 1.583,\ 1.561$ & $2.25\times10^{-2}$ \\
\bottomrule
\end{tabularx}
\caption{
Restricted branch crossings along the thermal-mass line.
CC and WW denote the intermediate color-block and weak-block branches, respectively.
The listed temperatures are the crossings between adjacent branches in the displayed sequence, and $\Delta T$ is defined in eq.~\eqref{eq:intermediate-width}.
}
\label{tab:thermal-crossings}
\end{table}

For $B_{\rm mild}$, $\beta+\gamma=0.02>0$, and a very narrow color-block interval appears between the $D_{32}$ and $H_{\rm sp}$ branches.
For $B_{\rm deep}$, $\beta+\gamma=-0.4<0$, and the sequence passes through a weak-block interval.
The two cases illustrate the orientation switch predicted by eq.~\eqref{eq:onset-splitting}.

\begin{figure}[t]
 \centering
 \includegraphics[width=0.98\linewidth]{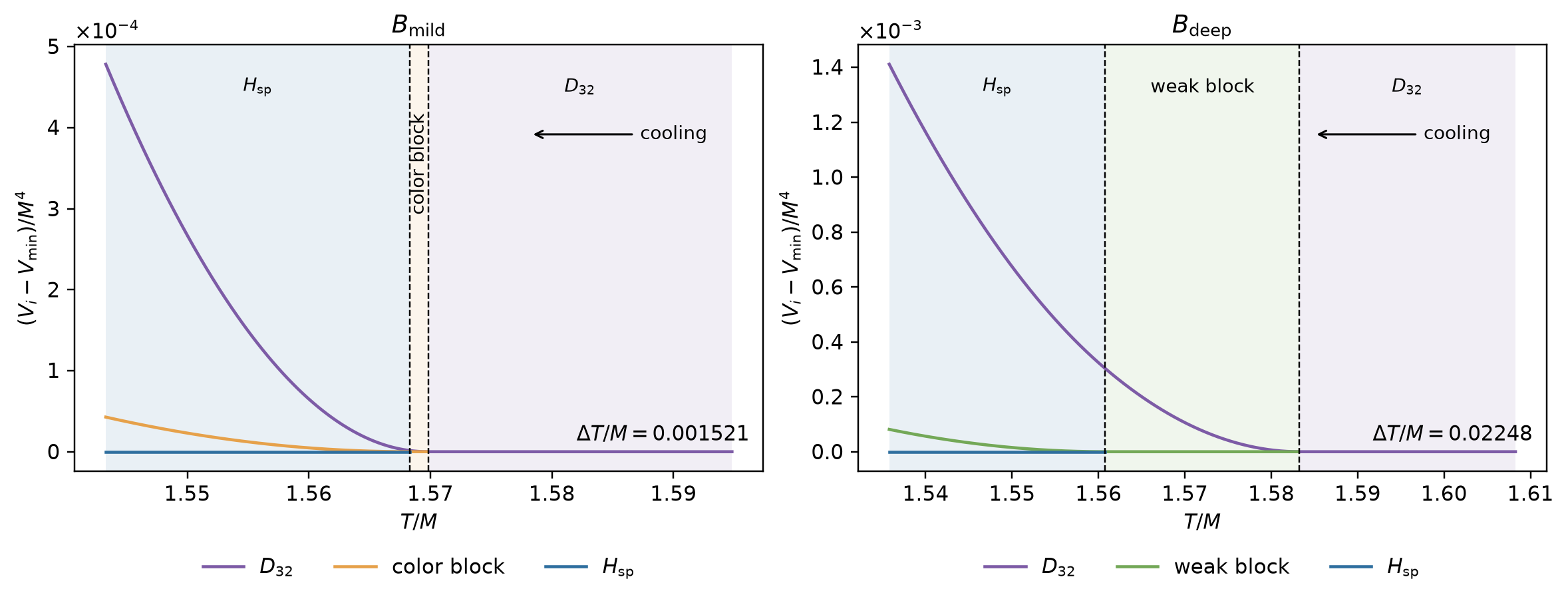}
 \caption{
 Restricted thermal branch ordering near the narrow intermediate intervals.
 Left: the color-block interval for $B_{\rm mild}$.
 Right: the weak-block interval for $B_{\rm deep}$.
 The shaded background denotes whether the $D_{32}$, intermediate, or $H_{\rm sp}$ branch is the lowest member of the comparison set; the curves show $V_i-V_{\min}$, and the dashed lines mark the refined crossings.
 The displayed $\Delta T/M$ is the separation between the two crossings that delimit the intermediate interval.
 Temperature increases to the right and the arrow indicates the cooling direction.
 }
 \label{fig:thermal-sequences}
\end{figure}

To display the energy structure near the crossings, we track
\begin{equation}
 \Delta V_i(T)=V_i(T)-V_{\min}(T).
\end{equation}
In figure~\ref{fig:thermal-sequences}, the shaded background identifies the lowest branch along the thermal line, while the curves show the branch energy differences.
Temperature increases to the right, so cooling proceeds from right to left.

The full 54-dimensional gradient and Hessian were also evaluated at the midpoint $T_{\rm mid}=(T_{\rm in}+T_{\rm out})/2$ of each intermediate interval.
The results are shown in table~\ref{tab:intermediate-stability}.

\begin{table}[t]
\centering
\begin{tabular}{lcccc}
\toprule
benchmark & intermediate branch & $\Delta T/M$ & $n_-^{\rm phys}$ & $m_{\min,+}^2/M^2$ \\
\midrule
$B_{\rm mild}$ & color block & $1.52\times10^{-3}$ & $0$ & $3.47\times10^{-4}$ \\
$B_{\rm deep}$ & weak block & $2.25\times10^{-2}$ & $0$ & $6.70\times10^{-3}$ \\
\bottomrule
\end{tabular}
\caption{
Full-field local stability at the midpoint of each intermediate interval.
Here $\Delta T$ is defined in eq.~\eqref{eq:intermediate-width}, $n_-^{\rm phys}$ is the number of negative eigenvalues after projecting out the symmetry tangent space, and $m_{\min,+}^2$ is the smallest strictly positive eigenvalue of the projected Hessian.
}
\label{tab:intermediate-stability}
\end{table}

At these midpoint temperatures, both intermediate branches are genuine local minima of the unrestricted field space rather than artifacts of constraining the fields to an endpoint family.
Representative points in the $D_{32}$ and $H_{\rm sp}$ intervals were checked in the same way and also showed no negative physical modes.

\subsection{Scope of the thermal analysis}

Equation~\eqref{eq:thermal-masses} is a leading thermal-mass model.
It does not include the one-loop Coleman--Weinberg contribution, the full finite-temperature functions, ring resummation, bubble nucleation, or real-time dynamics \cite{Coleman:1973jx,Dolan:1973qd,Arnold:1992rz}.
Tables~\ref{tab:thermal-crossings} and \ref{tab:intermediate-stability} therefore describe the ordering and local stability of stationary branches, not transition rates or cosmological trapping probabilities.

We treat only the entry from the $H_{32}$ phase into the intermediate $\Hsp$ phase.
The singlet-induced return
\begin{equation}
 \Hsp\longrightarrow H_{32}\simeq G_{\rm SM}
\end{equation}
and its gravitational-wave phenomenology in the complete Hamada--Yamatsu model are beyond the scope of this paper.

\section{Conclusions}
\label{sec:conclusions}

We have studied the local stability of subgroup-preserving stationary configurations in the renormalizable $SU(5)$ $24_H\oplus15_H$ Higgs sector, retaining all three mixed quartic portals.
On the standard $D_{32}$ background, the configurations preserving $H_{\rm sp}=S(O(3)\times O(2))$ form the two-block family $S=\diag(s_C I_3,s_L I_2)$.
Portal interactions can select unequal color- and weak-block amplitudes without changing the unbroken gauge group, so the isotropic configuration $S\propto I_5$ is a special representative rather than the definition of the phase.

The central result is that two independent orientation-sensitive quartics coincide on this two-block field space but contribute differently to color--weak transverse fluctuations.
We derived the complete cross-block spectrum, consisting of two sixfold physical families.
At fixed $\sigma=\beta+\gamma$, the reduced potential, stationary configuration, and energy remain unchanged while a transverse curvature can cross zero. An explicit bounded-from-below pair realizes this distinction as a local minimum in one case and a saddle with six negative physical modes in the other.
Independent Hessian calculations in the full 54-dimensional real scalar field space also confirm locally stable portal-distorted benchmarks.
The restricted branch comparison used to select these points is not a global-minimum certificate.

Within the leading thermal-mass approximation, the examples show that color- or weak-block condensation can precede entry into the interior $H_{\rm sp}$ family; the widths of the resulting single-block intervals are benchmark dependent.
The high-scale remnant required for Langacker--Pi monopole erasure can therefore be realized by locally stable, bounded-from-below portal-distorted configurations.
More generally, when distinct invariant interactions coincide on a symmetry-fixed field subspace, coupling information lost by the reduced potential can remain visible in the normal Hessian.
Full local stability therefore requires an explicit transverse test.

\acknowledgments
The author thanks Natsumi Nagata for a careful reading of the manuscript and for valuable comments and suggestions.
The author is also grateful to Yu Hamada and Naoki Yamatsu for reading the manuscript and for their kind and encouraging comments.

\appendix

\section{Derivation of the orbit-variable ranges}
\label{app:orbit}

\subsection{Adjoint invariant $u$}

Normalize $\Tr\Phi^2=1$ and denote the eigenvalues by $\phi_i$. The constraints are
\begin{equation}
 \sum_{i=1}^5\phi_i=0,
 \qquad
 \sum_{i=1}^5\phi_i^2=1,
\end{equation}
and $u=\sum_i\phi_i^4$.
Introducing Lagrange multipliers $\mu$ and $\nu_0$ for the unit-norm and tracelessness constraints, respectively, gives
\begin{equation}
 4\phi_i^3-2\mu\phi_i-\nu_0=0.
\end{equation}
Thus, a stationary configuration has at most three distinct eigenvalues.

For a two-valued configuration with multiplicities $\kappa$ and $5-\kappa$, the constraints give
\begin{equation}
 u_\kappa=\frac{(5-\kappa)^3+\kappa^3}{25\kappa(5-\kappa)}.
\end{equation}
The cases $\kappa=2,3$ give $u=7/30$, while $\kappa=1,4$ give $u=13/20$.
If the cubic equation has three distinct roots $a,b,c$, its vanishing quadratic coefficient implies $a+b+c=0$.
Up to permutation, the possible multiplicities are $(3,1,1)$ and $(2,2,1)$, giving $u=1/2$ and $u=1/4$, respectively.
The global extrema on the compact constraint surface are therefore the two-valued configurations,
\begin{equation}
 \frac{7}{30}\le u\le\frac{13}{20}.
\end{equation}
The lower and upper endpoints are realized by $D_{32}$ and $D_{41}$.

\subsection{Symmetric-tensor invariant $v$}

By Takagi factorization,
\begin{equation}
 S=U\diag(d_1,\ldots,d_5)U^T,
 \qquad d_i\ge0.
\end{equation}
After normalizing $\sum_i d_i^2=1$,
\begin{equation}
 v=\sum_i d_i^4.
\end{equation}
The standard simplex bounds give
\begin{equation}
 \frac15\le v\le1,
\end{equation}
with the lower endpoint at equal singular values and the upper endpoint at rank one.

\subsection{Two-block off-diagonal family and the CW balanced branch}
\label{app:CW}

Consider the normalized two-block adjoint direction
\begin{equation}
 D=\diag(aI_\kappa,bI_{5-\kappa}),
 \qquad
 \Tr D=0,
 \qquad
 \Tr D^2=1.
\end{equation}
These conditions imply
\begin{equation}
 a=\sqrt{\frac{5-\kappa}{5\kappa}},
 \qquad
 b=-\sqrt{\frac{\kappa}{5(5-\kappa)}},
 \qquad
 ab=-\frac15.
\end{equation}
If $S$ has support only in the off-diagonal block connecting the two eigenspaces,
\begin{equation}
 S=\begin{pmatrix}0&X\\X^T&0\end{pmatrix},
\end{equation}
then $DSD=abS$. With $N=\Tr(XX^\dagger)$, one has $\Tr(SS^\dagger)=2N$ and therefore
\begin{equation}
 q=ab=-\frac15,
 \qquad
 p=\frac{a^2+b^2}{2}.
\end{equation}
For $D_{32}$, $a=2/\sqrt{30}$ and $b=-3/\sqrt{30}$, so
\begin{equation}
 p=\frac{13}{60},
 \qquad
 q=-\frac15.
\end{equation}
If the two nonzero singular values of $X$ are $\sigma_1$ and $\sigma_2$, then
\begin{equation}
 v=\frac{\sigma_1^4+\sigma_2^4}{2(\sigma_1^2+\sigma_2^2)^2}.
\end{equation}
The balanced configuration $\sigma_1=\sigma_2$ has $v=1/4$, yielding
\begin{equation}
 (p,q,v)_{\rm CW,bal}=\left(\frac{13}{60},-\frac15,\frac14\right).
\end{equation}
The value $q=-1/5$ is specific to this two-block off-diagonal family and is not a global lower bound for general $(\Phi,S)$.

\section{Derivation of the color--weak cross-block masses}
\label{app:transverse}

Let $E_{ij}$ denote the standard matrix unit,
\begin{equation}
 (E_{ij})_{kl}=\delta_{ik}\delta_{jl}.
\end{equation}
For one color--weak pair $(a,\alpha)$, with $a=1,2,3$ and $\alpha=4,5$, we introduce two real fluctuation sectors.
\begin{align}
 e_\Phi^{(+)}&=\frac{i}{\sqrt2}(E_{a\alpha}-E_{\alpha a}),
 &
 e_S^{(+)}&=\frac{i}{\sqrt2}(E_{a\alpha}+E_{\alpha a}),
 \\
 e_\Phi^{(-)}&=\frac{1}{\sqrt2}(E_{a\alpha}+E_{\alpha a}),
 &
 e_S^{(-)}&=\frac{1}{\sqrt2}(E_{a\alpha}+E_{\alpha a}).
\end{align}
The repeated matrix in the second line belongs to two different field spaces.
We write
\begin{equation}
 \delta\Phi=x_\eta e_\Phi^{(\eta)},
 \qquad
 \delta S=y_\eta e_S^{(\eta)},
 \qquad \eta=\pm1.
\end{equation}
Each basis satisfies
\begin{equation}
 \Tr[(e_\Phi^{(\eta)})^2]=1,
 \qquad
 \Tr(e_S^{(\eta)}e_S^{(\eta)\dagger})=1.
\end{equation}
The difference between the color and weak adjoint eigenvalues is
\begin{equation}
 \Delta_\Phi=\phi_C-\phi_L=\frac{5\phi}{\sqrt{30}}=\sqrt{\frac56}\,\phi.
\end{equation}

The corresponding broken generators determine the gauge directions within these two-dimensional pair spaces.
The Hermitian generators
\begin{equation}
 T_{a\alpha}^{(+)}=\frac1{\sqrt2}(E_{a\alpha}+E_{\alpha a}),
 \qquad
 T_{a\alpha}^{(-)}=\frac{i}{\sqrt2}(E_{a\alpha}-E_{\alpha a})
\end{equation}
span the two broken real directions for each color--weak pair.
Their actions on the background are
\begin{align}
 i[T_{a\alpha}^{(+)},\Phi_0]&=-\Delta_\Phi e_\Phi^{(+)},
 &
 i(T_{a\alpha}^{(+)}S_0+S_0T_{a\alpha}^{(+)T})&=(s_C+s_L)e_S^{(+)},
 \\
 i[T_{a\alpha}^{(-)},\Phi_0]&=+\Delta_\Phi e_\Phi^{(-)},
 &
 i(T_{a\alpha}^{(-)}S_0+S_0T_{a\alpha}^{(-)T})&=(s_C-s_L)e_S^{(-)}.
\end{align}
Thus the gauge tangents are
\begin{equation}
 g_+=\begin{pmatrix}-\Delta_\Phi\\s_C+s_L\end{pmatrix},
 \qquad
 g_-=\begin{pmatrix}+\Delta_\Phi\\s_C-s_L\end{pmatrix}.
\end{equation}
Gauge invariance at a stationary point implies that the corresponding pair Hessians annihilate these vectors.
After applying the stationary-point conditions, they take the rank-one forms
\begin{equation}
 H_+=\frac{\Xi_+}{6}
 \begin{pmatrix}r_+^2&r_+\\r_+&1\end{pmatrix},
 \qquad
 r_+=\frac{s_C+s_L}{\Delta_\Phi},
\end{equation}
\begin{equation}
 H_-=\frac{\Xi_-}{6}
 \begin{pmatrix}r_-^2&-r_-\\-r_-&1\end{pmatrix},
 \qquad
 r_-=\frac{s_C-s_L}{\Delta_\Phi},
\end{equation}
where
\begin{equation}
 \Xi_+=12\kappa_2(s_C-s_L)^2-5\gamma\phi^2,
 \qquad
 \Xi_-=12\kappa_2(s_C+s_L)^2-5\gamma\phi^2.
\end{equation}
Each pair block contains one gauge rotation and one physical relative orientation.

To extract the physical eigenvalue, one must also account for the relative normalization of the adjoint and symmetric-tensor fluctuations in the kinetic term.
For either sector, substitution into the kinetic term gives
\begin{equation}
 \mathcal L_{\rm kin}^{(2)}
 =\frac12(\partial x_\eta)^2+(\partial y_\eta)^2
 =\frac12(\partial_\mu z_\eta)^TG(\partial^\mu z_\eta),
 \qquad
 G=\begin{pmatrix}1&0\\0&2\end{pmatrix},
\end{equation}
where $z_\eta=(x_\eta,y_\eta)^T$.
The physical masses are the generalized eigenvalues
\begin{equation}
 H_\eta u=m^2Gu.
\end{equation}
Each block has one zero eigenvalue and one nonzero eigenvalue,
\begin{equation}
 m_{\perp,\eta}^2
 =\frac{\Xi_\eta}{6}\left(r_\eta^2+\frac12\right)
 =\left[\frac1{12}+\frac{(s_C+\eta s_L)^2}{5\phi^2}\right]\Xi_\eta,
 \qquad \eta=\pm1.
\end{equation}
Each real sector occurs for all six color--weak pairs, producing two sixfold physical families.

\section{A bounded-from-below minimum--saddle pair}
\label{app:stable-saddle-pair}

On the reference slice in eq.~\eqref{eq:benchmark-slice}, at $\widehat\rho=0$ and fixed $\sigma=\beta+\gamma=0.02$, the two points in table~\ref{tab:transverse-boundary-pair} have the same reduced potential.
They therefore share
\begin{equation}
 \phi=0.7634,\quad
 s_C=0.3469,\quad
 s_L=0.3455,\quad
 \frac{V}{M^4}=-0.8869.
 \label{eq:shared-pair-background}
\end{equation}

\begin{table}[t]
\centering
\begin{tabular}{lccccc}
\toprule
point & $\beta$ & $\gamma$ & $m_{\perp,+}^2/M^2$ & $m_{\perp,-}^2/M^2$ & $n_-^{\rm phys}$ \\
\midrule
$P_{\rm min}$ & $0.100$ & $-0.080$ & $0.05779$ & $0.4988$ & $0$ \\
$P_{\rm sad}$ & $0.019$ & $0.001$ & $-0.0007164$ & $0.4792$ & $6$ \\
\bottomrule
\end{tabular}
\caption{
A bounded-from-below minimum--saddle pair at fixed $\sigma=\beta+\gamma=0.02$.
The reduced potential, stationary VEV, and vacuum energy are identical, while the $\eta=+1$ cross-block family changes sign. $n_-^{\rm phys}$ counts negative physical eigenvalues of the full 54-dimensional Hessian.
}
\label{tab:transverse-boundary-pair}
\end{table}

Both potentials are bounded from below.
For $P_{\rm min}$ this follows from the analytic bound in section~\ref{sec:bfb-benchmarks}.
For $P_{\rm sad}$, the companion identity
\begin{equation}
 \Tr(\Phi^2SS^\dagger)+\Tr(\Phi S\Phi^T S^\dagger)
 =\frac12\Tr\!\left[(\Phi S+S\Phi^T)(\Phi S+S\Phi^T)^\dagger\right]
 \ge0
 \label{eq:pplusq-raw}
\end{equation}
combines with eq.~\eqref{eq:pminusq-raw} to give $|q|\le p$.
Since $\beta\ge\gamma\ge0$ at $P_{\rm sad}$,
\begin{equation}
 C(p,q)=\alpha+\beta p+\gamma q
 \ge\alpha+(\beta-\gamma)p
 \ge\alpha>0,
 \label{eq:saddle-pair-bfb}
\end{equation}
while the pure-sector margins are unchanged.
The saddle is therefore caused by a genuine transverse instability, not by an unbounded quartic direction.

This pair makes the transverse local-stability statement operational: the restricted theory cannot distinguish $P_{\rm min}$ from $P_{\rm sad}$, whereas the complete theory assigns them different local stability: one is a minimum and the other a saddle.

\bibliographystyle{JHEP}
\bibliography{references}

@article{Georgi:1974sy,
    author = "Georgi, H. and Glashow, S. L.",
    title = "{Unity of All Elementary Particle Forces}",
    doi = "10.1103/PhysRevLett.32.438",
    journal = "Phys. Rev. Lett.",
    volume = "32",
    pages = "438--441",
    year = "1974"
}

@article{tHooft:1974kcl,
    author = "'t Hooft, Gerard",
    editor = "Taylor, J. C.",
    title = "{Magnetic Monopoles in Unified Gauge Theories}",
    reportNumber = "CERN-TH-1876",
    doi = "10.1016/0550-3213(74)90486-6",
    journal = "Nucl. Phys. B",
    volume = "79",
    pages = "276--284",
    year = "1974"
}

@article{Polyakov:1974ek,
    author = "Polyakov, Alexander M.",
    editor = "Taylor, J. C.",
    title = "{Particle Spectrum in Quantum Field Theory}",
    reportNumber = "PRINT-74-1566 (LANDAU-INST)",
    journal = "JETP Lett.",
    volume = "20",
    pages = "194--195",
    year = "1974"
}

@article{Kibble:1976sj,
    author = "Kibble, T. W. B.",
    title = "{Topology of Cosmic Domains and Strings}",
    reportNumber = "ICTP/75/5",
    doi = "10.1088/0305-4470/9/8/029",
    journal = "J. Phys. A",
    volume = "9",
    pages = "1387--1398",
    year = "1976"
}

@article{Preskill:1979zi,
    author = "Preskill, John",
    title = "{Cosmological Production of Superheavy Magnetic Monopoles}",
    reportNumber = "HUTP-79/A028",
    doi = "10.1103/PhysRevLett.43.1365",
    journal = "Phys. Rev. Lett.",
    volume = "43",
    pages = "1365",
    year = "1979"
}

@article{Langacker:1980kd,
    author = "Langacker, Paul and Pi, So-Young",
    title = "{Magnetic Monopoles in Grand Unified Theories}",
    reportNumber = "SLAC-PUB-2496, Print-80-0262 (IAS,PRINCETON)",
    doi = "10.1103/PhysRevLett.45.1",
    journal = "Phys. Rev. Lett.",
    volume = "45",
    pages = "1",
    year = "1980"
}

@article{Palais:1979rca,
    author = "Palais, Richard S.",
    title = "{The principle of symmetric criticality}",
    doi = "10.1007/BF01941322",
    journal = "Commun. Math. Phys.",
    volume = "69",
    number = "1",
    pages = "19--30",
    year = "1979"
}

@article{Hamada:2026iht,
    author = "Hamada, Yu and Yamatsu, Naoki",
    title = "{A simple solution to the monopole problem: SU(5) GUT with symmetry breaking into special subgroup}",
    eprint = "2606.12874",
    archivePrefix = "arXiv",
    primaryClass = "hep-ph",
    reportNumber = "YITP-26-53, OU-HET-1307",
    month = "6",
    year = "2026"
}

@article{Frautschi:1981jh,
    author = "Frautschi, Steven C. and Kim, Jaisam",
    title = "{SU(5) Higgs Problem With Adjoint + Vector Representations}",
    reportNumber = "CALT-68-859",
    doi = "10.1016/0550-3213(82)90041-4",
    journal = "Nucl. Phys. B",
    volume = "196",
    pages = "301--327",
    year = "1982"
}

@article{Sartori:1981zj,
    author = "Sartori, G.",
    title = "{A Theorem on Orbit Structures (Strata) of Compact Lie Groups}",
    reportNumber = "IFPD 65/81",
    doi = "10.1063/1.525772",
    journal = "J. Math. Phys.",
    volume = "24",
    pages = "765",
    year = "1983"
}

@inproceedings{Talamini:2006cf,
    author = "Talamini, Vittorino",
    title = "{Orbit spaces of compact linear groups}",
    booktitle = "{4th International School on Theoretical Physics: Symmetry and Structural Properties of Condensed Matter (SSPCM)}",
    eprint = "hep-th/0607245",
    archivePrefix = "arXiv",
    month = "7",
    year = "2006"
}

@article{Slansky:1981yr,
    author = "Slansky, R.",
    title = "{Group Theory for Unified Model Building}",
    reportNumber = "LA-UR-80-3495",
    doi = "10.1016/0370-1573(81)90092-2",
    journal = "Phys. Rept.",
    volume = "79",
    pages = "1--128",
    year = "1981"
}

@article{Kannike:2012pe,
    author = "Kannike, Kristjan",
    title = "{Vacuum Stability Conditions From Copositivity Criteria}",
    eprint = "1205.3781",
    archivePrefix = "arXiv",
    primaryClass = "hep-ph",
    doi = "10.1140/epjc/s10052-012-2093-z",
    journal = "Eur. Phys. J. C",
    volume = "72",
    pages = "2093",
    year = "2012"
}

@article{Dolan:1973qd,
    author = "Dolan, L. and Jackiw, R.",
    title = "{Symmetry Behavior at Finite Temperature}",
    reportNumber = "MIT-CTP-406",
    doi = "10.1103/PhysRevD.9.3320",
    journal = "Phys. Rev. D",
    volume = "9",
    pages = "3320--3341",
    year = "1974"
}

@article{Coleman:1973jx,
    author = "Coleman, Sidney R. and Weinberg, Erick J.",
    title = "{Radiative Corrections as the Origin of Spontaneous Symmetry Breaking}",
    doi = "10.1103/PhysRevD.7.1888",
    journal = "Phys. Rev. D",
    volume = "7",
    pages = "1888--1910",
    year = "1973"
}

@article{Arnold:1992rz,
    author = "Arnold, Peter Brockway and Espinosa, Olivier",
    title = "{The Effective potential and first order phase transitions: Beyond leading-order}",
    eprint = "hep-ph/9212235",
    archivePrefix = "arXiv",
    reportNumber = "UW-PT-92-18, USM-TH-60",
    doi = "10.1103/PhysRevD.47.3546",
    journal = "Phys. Rev. D",
    volume = "47",
    pages = "3546",
    year = "1993",
    note = "[Erratum: Phys.Rev.D 50, 6662 (1994)]"
}

@article{Abud:1981tf,
    author = "Abud, M. and Sartori, G.",
    title = "{The Geometry of Orbit Space and Natural Minima of Higgs Potentials}",
    reportNumber = "IFPD 45/81",
    doi = "10.1016/0370-2693(81)90578-5",
    journal = "Phys. Lett. B",
    volume = "104",
    pages = "147--152",
    year = "1981"
}

@article{Kim:1981xu,
    author = "Kim, Jaisam",
    title = "{General Method for Analyzing Higgs Potentials}",
    reportNumber = "CALT-68-857",
    doi = "10.1016/0550-3213(82)90040-2",
    journal = "Nucl. Phys. B",
    volume = "196",
    pages = "285--300",
    year = "1982"
}

@article{Abud:1983id,
    author = "Abud, M. and Sartori, G.",
    title = "{The Geometry of Spontaneous Symmetry Breaking}",
    reportNumber = "Print-83-0073 (PADUA)",
    doi = "10.1016/0003-4916(83)90017-9",
    journal = "Annals Phys.",
    volume = "150",
    pages = "307",
    year = "1983"
}

@article{Ruegg:1980gf,
    author = "Ruegg, H.",
    title = "{Extremas of SU($N$) Higgs Potentials and Symmetry Breaking Pattern}",
    reportNumber = "SLAC-PUB-2518",
    doi = "10.1103/PhysRevD.22.2040",
    journal = "Phys. Rev. D",
    volume = "22",
    pages = "2040",
    year = "1980"
}

@article{Kim:1981jj,
    author = "Kim, Jaisam",
    title = "{SU($N$) Higgs Problem With Adjoint Representation and Michel's Conjecture}",
    reportNumber = "CALT-68-864",
    doi = "10.1016/0550-3213(82)90160-2",
    journal = "Nucl. Phys. B",
    volume = "197",
    pages = "174--188",
    year = "1982"
}

@article{Chakrabortty:2013mha,
    author = "Chakrabortty, Joydeep and Konar, Partha and Mondal, Tanmoy",
    title = "{Copositive Criteria and Boundedness of the Scalar Potential}",
    eprint = "1311.5666",
    archivePrefix = "arXiv",
    primaryClass = "hep-ph",
    doi = "10.1103/PhysRevD.89.095008",
    journal = "Phys. Rev. D",
    volume = "89",
    number = "9",
    pages = "095008",
    year = "2014"
}

@article{Kannike:2016fmd,
    author = "Kannike, Kristjan",
    title = "{Vacuum Stability of a General Scalar Potential of a Few Fields}",
    eprint = "1603.02680",
    archivePrefix = "arXiv",
    primaryClass = "hep-ph",
    doi = "10.1140/epjc/s10052-016-4160-3",
    journal = "Eur. Phys. J. C",
    volume = "76",
    number = "6",
    pages = "324",
    year = "2016",
    note = "[Erratum: Eur.Phys.J.C 78, 355 (2018)]"
}

@article{Chauhan:2019fji,
    author = "Chauhan, Garv",
    title = "{Vacuum Stability and Symmetry Breaking in Left-Right Symmetric Model}",
    eprint = "1907.07153",
    archivePrefix = "arXiv",
    primaryClass = "hep-ph",
    doi = "10.1007/JHEP12(2019)137",
    journal = "JHEP",
    volume = "12",
    pages = "137",
    year = "2019"
}

\end{document}